\documentclass[12pt,preprint]{aastex}
\usepackage{epsfig}

\def\la{\mathrel{\mathpalette\fun <}}

\def\fun#1#2{\lower3.6pt\vbox{\baselineskip0pt\lineskip.9pt
  \ialign{$\mathsurround=0pt#1\hfil##\hfil$\crcr#2\crcr\sim\crcr}}}

\newcommand{\um}{12}
\newcommand{\cit}{1}
\newcommand{\lbl}{2}
\newcommand{\stockholm}{5}
\newcommand{\lpnhe}{6}
\newcommand{\yale}{9}
\newcommand{\lam}{8}
\newcommand{\upenn}{10}
\newcommand{\ucb}{11}
\newcommand{\stsci}{12}
\newcommand{\cppm}{7}
\newcommand{\iu}{13}
\newcommand{\aas}{16}
\newcommand{\uc}{17}
\newcommand{\cambridge}{18}
\newcommand{\cea}{19}
\newcommand{\ipnl}{20}
\newcommand{\slac}{4}
\newcommand{\fnal}{3}
\newcommand{\anu}{15}
\newcommand{\jpl}{14}

\begin{document} 

\title{Seeing the Nature of the Accelerating Physics: It's a SNAP\\ 
{\it Overview White Paper to the Dark Energy Task Force}} 

\renewcommand{\thefootnote}{\fnsymbol{footnote}}
\author{J.~Albert\altaffilmark{\cit},
G.~Aldering\altaffilmark{\lbl},
S.~Allam\altaffilmark{\fnal},
W.~Althouse\altaffilmark{\slac},
R.~Amanullah\altaffilmark{\stockholm}, J.~Annis\altaffilmark{\fnal},
P.~Astier\altaffilmark{\lpnhe},
M.~Aumeunier\altaffilmark{\cppm,\lam},
S.~Bailey\altaffilmark{\lbl},
C.~Baltay\altaffilmark{\yale}
% ...
}

\email{saul@lbl.gov, melevi@lbl.gov}

\altaffiltext{\cit}{California Institute of Technology}
\altaffiltext{\lbl}{Lawrence Berkeley National Laboratory}
\altaffiltext{\fnal}{Fermi National Accelerator Laboratory}
\altaffiltext{\slac} {Stanford Linear Accelerator Center}
\altaffiltext{\stockholm}{University of Stockholm}
\altaffiltext{\lpnhe}{LPNHE, CNRS-IN2P3, Paris, France}
\altaffiltext{\cppm}{CPPM, CNRS-IN2P3, Marseille, France}
\altaffiltext{\lam}{LAM, CNRS-INSU, Marseille, France}
\altaffiltext{\yale}{Yale University}
\altaffiltext{\upenn}{University of Pennsylvania}
\altaffiltext{\ucb}{University of California at Berkeley}
\altaffiltext{\um}{University of Michigan}
\altaffiltext{\stsci}{Space Telescope Science Institute}
\altaffiltext{\iu}{Indiana University}
\altaffiltext{\jpl}{Jet Propulsion Laboratory}
\altaffiltext{\anu}{The Australian National University}
\altaffiltext{\aas}{American Astronomical Society}
\altaffiltext{\uc}{University of Chicago}
\altaffiltext{\cambridge}{Cambridge University}
\altaffiltext{\cea}{CEA, Saclay, France}
\altaffiltext{\ipnl}{IPNL, CNRS-IN2P3, Villeurbanne, France}

\newpage 

\section{Summary} 

\begin{figure}[!h]
\begin{center} 
\psfig{file=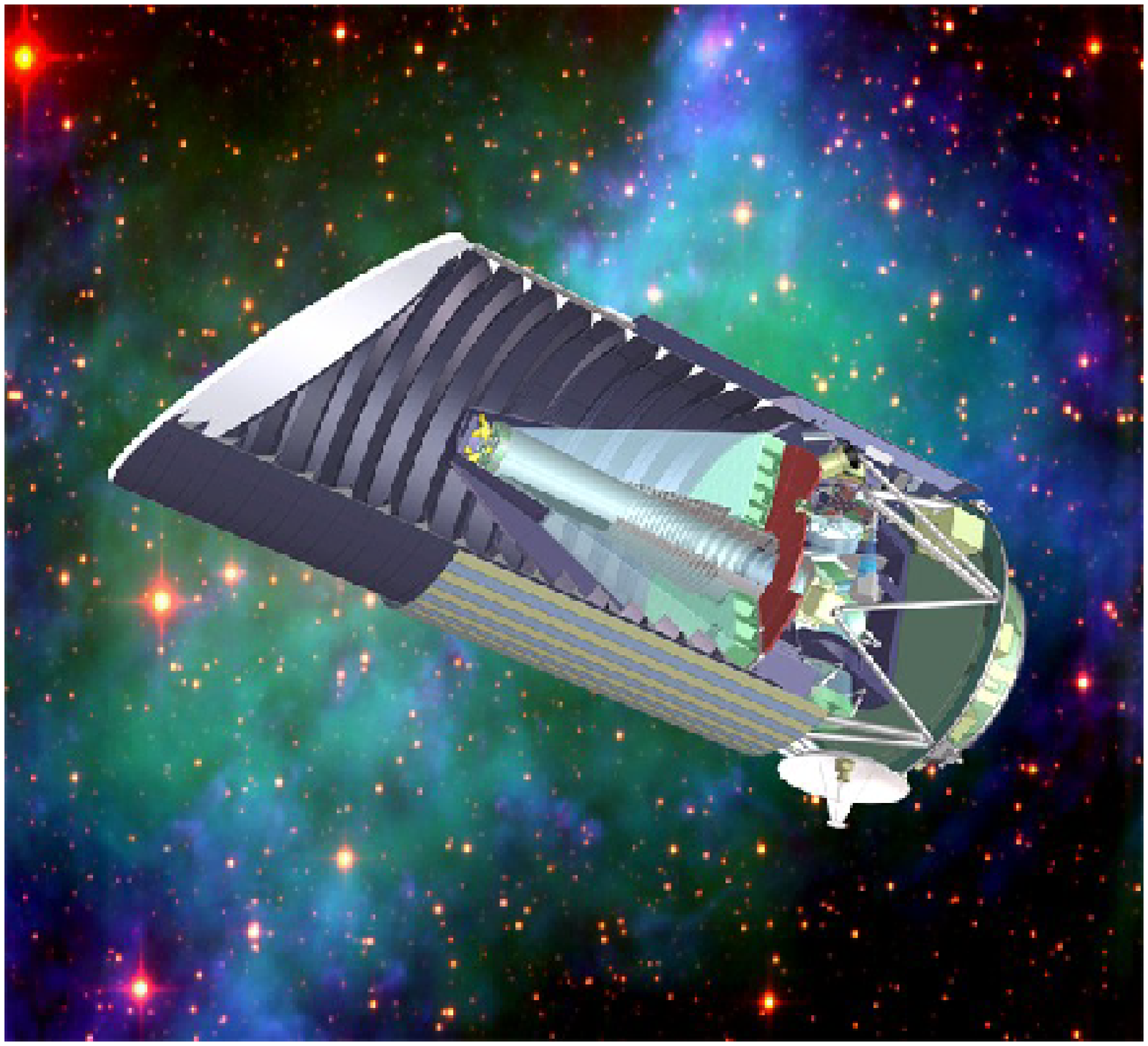,width=2.in} 
\end{center} 
\end{figure} 

For true insight into the nature of 
dark energy, 
measurements of the precision and accuracy of the Supernova/Acceleration 
Probe (SNAP) 
are required.  Precursor or scaled-down experiments are unavoidably 
limited, even 
for distinguishing the cosmological constant.  
They can pave the way for, but should not delay, SNAP by developing 
calibration, refinement, and systematics control (and they will also provide 
important, exciting astrophysics).

\vspace{0.2in} 

\begin{enumerate} 

\item To understand dark energy requires accurate knowledge of the 
physical dynamics $w$, $w'$.   
\item To see accurately the dynamics requires mapping of the expansion 
history, or geometric distance measurements, covering the full range 
$z=0-1.7$. 
\item To achieve robust results requires stringent systematics controls 
within a cosmological technique and crosschecking between techniques. 
Complementarity also provides enhanced constraints.  It is particularly 
essential to conjoin a distance probe and a mass growth probe (e.g.\ 
supernovae and weak lensing) to reveal the physical origin of dark energy. 
\item To attain the first three points requires the third generation of 
experiments, i.e.\ the Joint Dark Energy Mission (JDEM), e.g.\ SNAP.  
Experiments without $w'$, the redshift range, 
mature, identified, and controlled systematics limits, and complementarity 
do not provide a substantial understanding of dark energy, despite 
statistical gains.  Precursor 
experiments add value in developing systematics control and cosmological 
techniques directly in line with the third generation. 

\end{enumerate}

\section{The Role of Precursors, The Role of JDEM} 

We need to meet the extraordinary challenge of the new physics behind 
cosmic acceleration by designing revelatory, robust experiments.  
Revelatory means detailed investigation of the dynamics of the 
dark energy, characterizing its equation of state value and variation, e.g.\ 
the ``tilt'' $1+w$ and ``running'' $w'=\dot w/H$.  Robust means that 
we must have confidence that the data tell us the true answer, not one 
biased or degraded due to systematic uncertainties. 

What we can do with JDEM is put together the most stringent, 
deep- and wide-seeing experiment possible in the next generation.  
A space mission requires strong justification and must 
provide a substantial step beyond previous experiments or what can be 
achieved from the ground:  this has driven the SNAP design. 
For further details, see the SNAP supernovae and weak lensing white papers 
to the Dark Energy Task Force. 

\subsection{Dynamics} 

\begin{figure}[!hbt]
\begin{center} 
\psfig{file=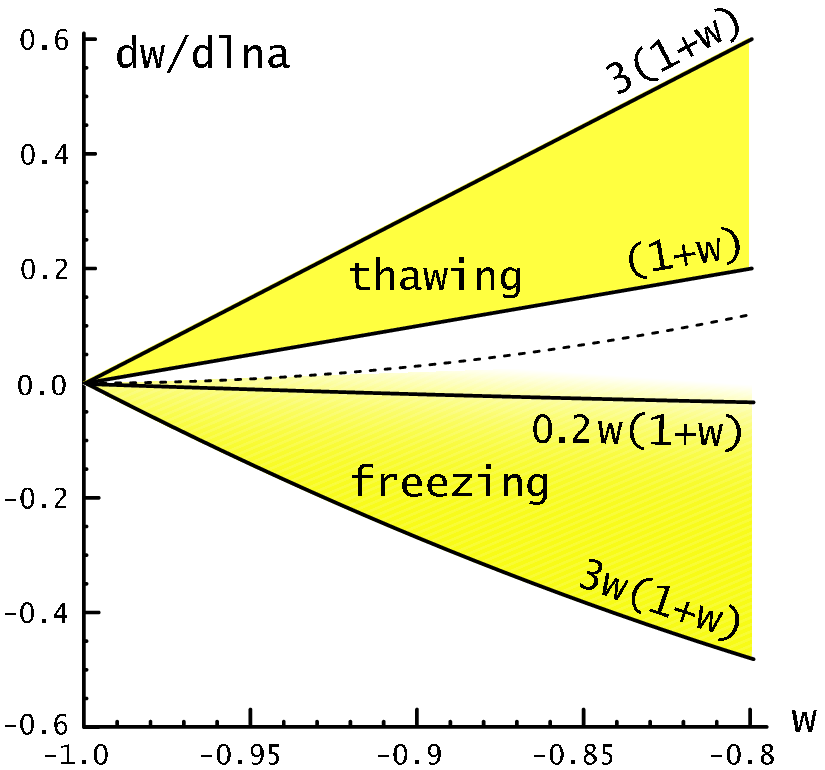,width=6.in} 
\caption{Scalar field models of dark energy can be separated into 
two distinct behaviors based on their dynamics, occupying narrow 
regions of the $w-w'$ phase space.  ``Freezing'' models initially 
roll and then slow to a creep as they come to dominate the Universe. 
``Thawing'' models initially are frozen and look like a cosmological 
constant, at $w=-1$, $w'=0$, 
and then thaw and roll to $w'>0$.  Despite this dynamics, all 
models in the thawing region above would be mistaken for a cosmological 
constant ($w=-1$) by a ground based experiment with 5\% precision on 
$w_{\rm constant}$.  From \citet{caldlin}. 
} 
\label{fig.shaded}
\end{center} 
\end{figure}

Precursors can give us a {\it glimpse\/} of dark energy in the form of 
$w_{\rm constant}$, but do not allow {\it study\/} of dark energy.  
Indeed, of the two major classes of quintessence models,  the half 
that are ``thawing'' 
models \citep{caldlin}, will appear {\it nearly identical\/} 
to a cosmological constant 
when viewed in terms of $w$ a priori constant, despite its dynamics 
(see Fig.~\ref{fig.shaded}). 
For example, an experiment aiming at a precision of 0.05 in 
$w_{\rm constant}$ would think it has found $w=-1\pm0.05$, concealing 
a thawing model with $w_0=-0.8$.  Such a biased result mistakes the 
physics for half the models in the phase space.   Thus $w'$ is essential, 
even for 
the question of whether dark energy is the cosmological constant or not. 

As a true next generation experiment, SNAP will give strong 
constraints on $w$ and $w'$ and guide us in the 
quest for the nature of the new physics.  Precursor experiments cannot match 
this fundamental requirement, being basically blind to the dynamics $w'$ on 
scales finer than the Hubble time \citep{linmiq}.  Indeed a 
basic physics 
distinction in scalar field physics requires the precision on the dynamics 
to be of order the deviation of the equation of state from the cosmological 
constant value, $\sigma(w')\sim 2(1+w)\la 0.1$ \citep{caldlin}. 

The role of precursor experiments then is not fundamentally one of revelation, 
but of robustness.  This does not stop them from doing exciting astrophysical 
science, but for dark energy their main role should be to move full speed 
ahead toward enabling JDEM.  They should be valuable contributions to the 
development 
and refinement of experimental techniques, control of systematics, and 
astrophysical calibration.  These are essential roles. 

A wide field telescope in space has 
access to a wide variety of cosmological probes, depending on survey strategy. 
These include Type Ia supernovae, weak gravitational lensing, baryon acoustic 
oscillations, strong gravitational lensing, Type II supernovae, and cluster 
properties and abundances.  Precursor experiments have an opportunity 
to realize, understand, and refine these varied techniques.  
During this development, the areas of advantage and disadvantage of 
each probe will become clearer, and systematic uncertainties must be 
identified and strategies designed to control them, before they can 
be considered seriously as useful tools. 
However, only SN Ia and weak lensing are developed 
to the point of currently being on the playing field of cosmological 
usefulness, and only SN Ia are mature.  

Concurrently, the unprecedented depth and precision of SNAP 
must be leveraged by a firm foundation of calibration -- 
understanding of photometric zeropoints, creation of standard star 
networks, crosswavelength calibration, PSF and atmospheric corrections, 
photometric redshift fitting, understanding heterogeneity of probe objects, 
etc.  As one example, extremely well calibrated low redshift supernovae 
studies (see, e.g., the spectrophotometric approach of 
the Nearby Supernova Factory \citep{woodvasey}) 
have always been treated as an essential component of cosmology fitting 
for the SN Ia distance method.  These precursor aspects are all 
true science and will further provide a lasting legacy for astrophysics. 

\subsection{Redshift Range} 

\begin{figure}[!hbt] 
\begin{center} 
\psfig{file=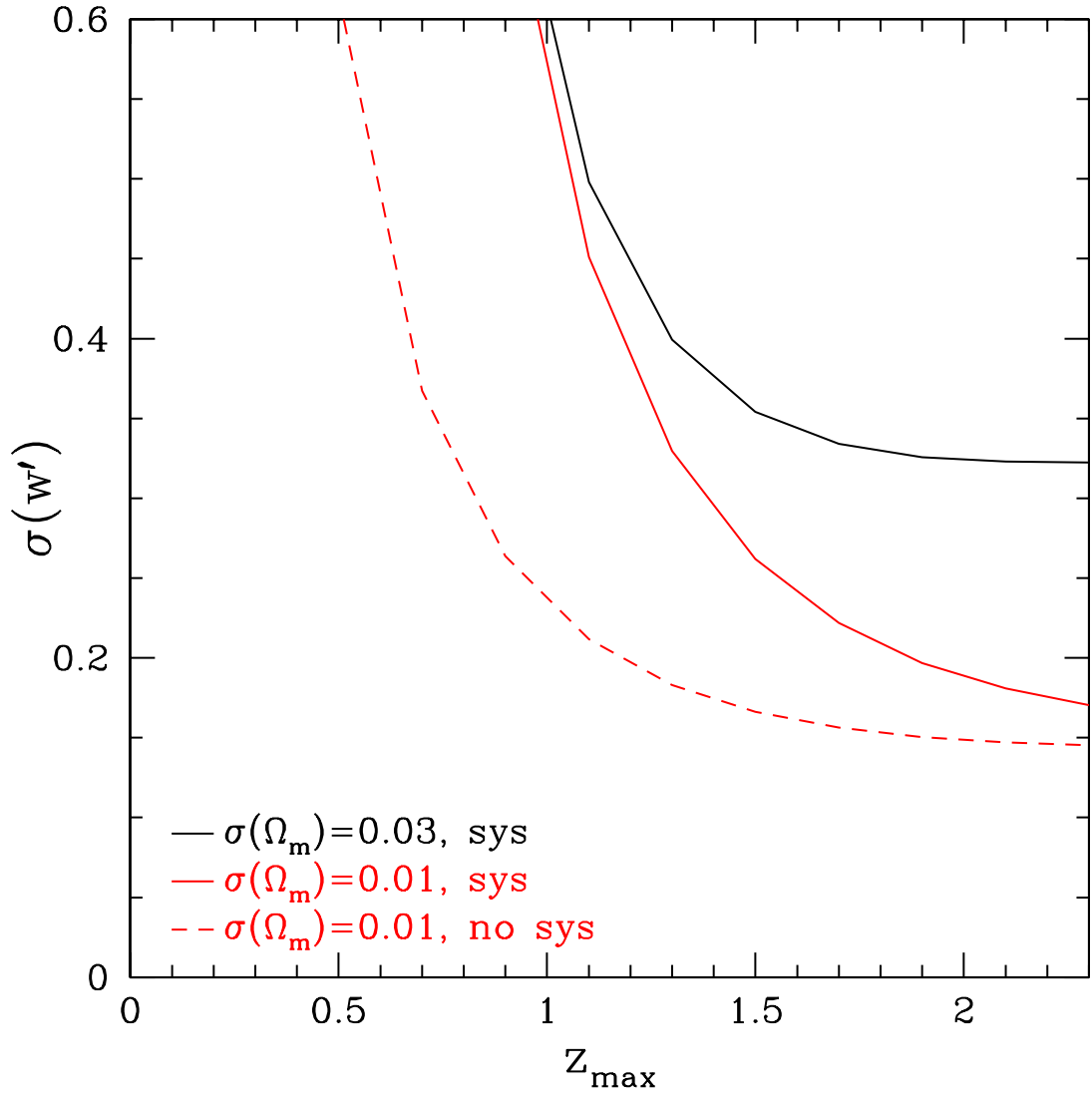,width=6.in} 
\caption{Uncertainty in determination of the time variation of the 
dark energy equation of state as a function of distance survey 
depth $z_{\rm max}$.  Even in the idealized case of no systematic 
error the uncertainty rises steeply as $z_{\rm max}$ decreases. 
To detect the key discriminator of fundamental physics one requires 
a survey extending to $z_{\rm max}>1.5$.  From \citet{linhut03}. 
} 
\label{fig.sigcw1} 
\end{center} 
\end{figure}

\begin{figure}[!hbt]
\begin{center}
\psfig{file=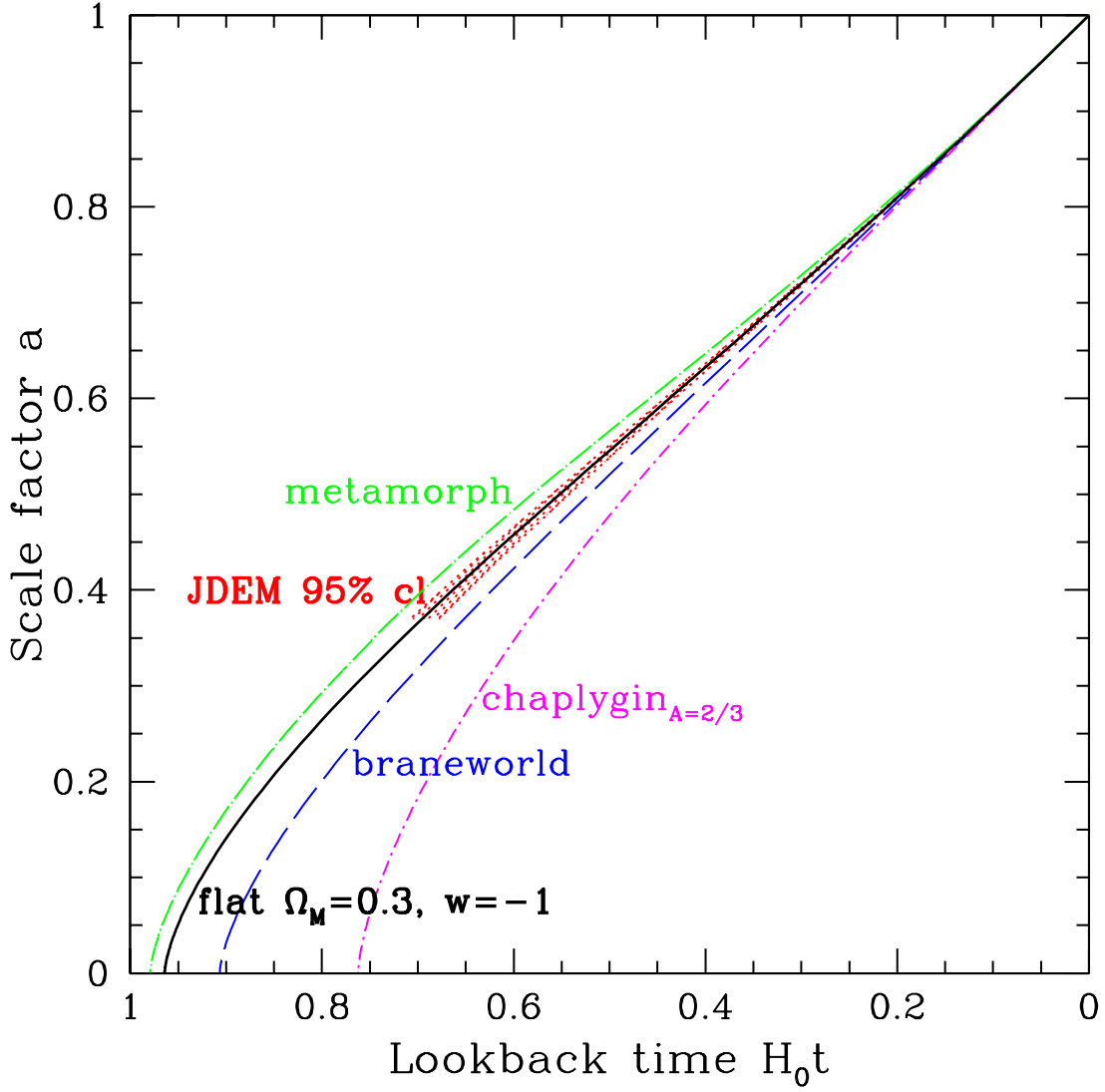,width=6.in} 
\caption{With percent level mapping of the cosmic expansion history 
by the direct supernova distance-redshift relation comes guidance to 
the nature of the dark energy, whether 
physics involving structure in the quantum vacuum (e.g.\ metamorph), 
extra-dimensional extensions to gravity 
(e.g.\ braneworld), interaction between dark energy and dark matter 
(e.g.\ chaplygin), or a fundamental cosmological constant $\Lambda$. 
Less accurate measurements will merely leave us confused. 
Knowledge of our history may then allow us to look to the future 
and explore the fate of the universe. 
}
\label{fig.his2}
\end{center}
\end{figure}

Redshift depth, completeness, and homogeneity of the sample are all 
key issues for both the leverage and robustness of the data.  For 
distance measurements, the sensitivity curve of 
determining either $w$ or $w'$ poses a steep obstacle 
at redshifts $z<1.5$ (see Fig.~\ref{fig.sigcw1}; \citet{linhut03}), 
so data sampling the entire region from low redshift to 
$z>1.5$ (with unified calibration) is necessary.  Supernovae give a 
direct, geometric probe of the expansion history of the universe, an 
essential tool for understanding dark energy and a landmark of cosmology 
in its own right.  Cobbling together disparate experiments will not 
provide the accuracy needed; indeed offsets of as little as 0.02 mag 
between redshift sets can lead to biases of order $0.7\sigma$ \citep{linmiq}. 
These requirements, however, are 
beyond the reach of precursor experiments, and even handfuls of 
measurements at $z>1$ cannot avoid systematic 
bias from gravitational lensing magnification and other observational 
difficulties.  A unified supernova distance experiment covering the 
full range $z=0.1-1.7$ is essential for understanding dark energy 
(see Fig.~\ref{fig.his2}).  

For weak gravitational lensing, depth is also a key advantage, 
strongly increasing the cosmological leverage, more so than sky area. 
The effective number density of galaxies entering the 
shear measurements scales as 
$$
n_{\rm eff}(z)\sim n_g C_l(z)/C_l(z_{\rm fid})\approx n_g[1+2.5(z_m-1)]/
[1+2.5(z_{\rm fid}-1)], 
$$ 
with $C_l$ is the shear power, $z_m$ is the median source redshift, and 
$z_{\rm fid}$ is the median source redshift for a comparison survey 
\citep{valelin}. 
This implies that a survey with $n_g=100/{\rm arcmin}^2$ and $z_m=1.2$ has 
not 3.3, but 10 times the effective number density as a survey with 
$n_g=30/{\rm arcmin}^2$ 
and $z_m=0.8$.  Such depth can therefore compensate for a {\it factor 100 
in sky area\/}, in the shape noise dominated regime.  SNAP will not 
have any competition in that regime from ground based weak lensing surveys 
before the Large Survey Telescope (LST). 

\subsection{Complementarity}  

\begin{figure}[!hbt]
\begin{center}
\psfig{file=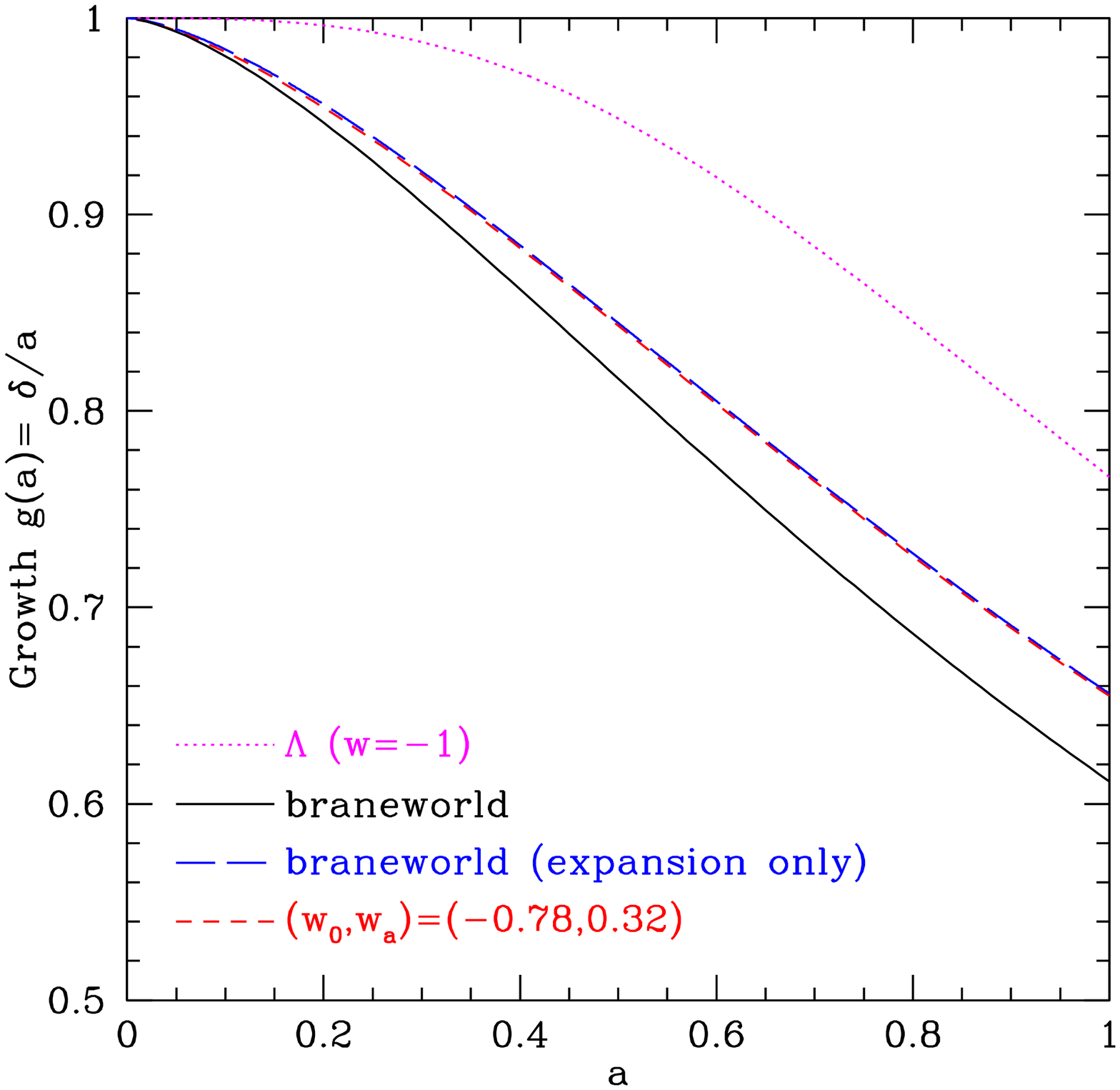,width=5.6in}
\caption{Growth history for linear perturbations in matter density is 
shown as a function of scale factor.  This provides an important 
window on dark energy distinct from expansion history, since it also 
reacts to gravitational modifications.  Extra-dimensional extension 
to gravity in the braneworld scenario modifies the growth so as to 
shift the long dashed, blue curve to the black, solid behavior, a 
deviation of 2-7\% for $z=0-2$.  (The cosmological constant case, 
dotted magenta, is shown for comparison.)  
Without accounting for this modification, a scalar field model with 
the same expansion history (short dashed, red curve) could not be 
distinguished from the braneworld model.  Conversely, a scalar field 
model agreeing with the growth history of the braneworld model would 
deviate in the expansion history.  Therefore, accurate measurements of 
both expansion history and growth history, e.g.\ supernovae and weak 
lensing, are required to understand dark energy. 
} 
\label{fig.bwgro} 
\end{center}
\end{figure}

Combining a purely geometric measure of distance, such as supernovae, 
with a probe that includes a sensitivity to mass growth, such as weak 
lensing opens important new avenues for understanding the physical 
origin of dark energy.  While the growth information in weak lensing 
(or any other such probe) is admixed with distances in a complicated 
fashion, in synergy with supernova measurements it offers 
the opportunity for distinguishing a high 
energy physics component from an extension to the theory of gravity 
as an explanation for the acceleration of the universe (see 
Fig.~\ref{fig.bwgro}; \citet{lin05a}), i.e.\ testing whether Einstein 
gravity breaks down.  Neither probe alone accomplish this. 

The essential development, refinement, and calibration 
of cosmological techniques, and voluminous increase in our astrophysical 
knowledge, produced by precursor experiments are real science, and 
furthermore provide 
crucial foundations to the next generation of JDEM and LST.  

\section{``Ensure Rapid Progress''} 

The key role of precursor experiments is laid out in the agency charge 
to the Dark Energy Task Force: ``ensure rapid progress... towards 
understanding the nature of dark energy''.  As shown above, that 
understanding will not come before SNAP.  Precursor experiments 
should aim to provide rapid progress in realizing the far more 
comprehensive next generation. 

The greatest leverage will come through increasing the robustness of 
the cosmological probes that have already proved themselves capable. 
Indeed, concrete, essential contributions have already been identified 
and are being implemented: 

\begin{enumerate} 

\item Detailed characterization of supernova heterogeneity through 
spectrophotometric study in a wide variety of environments.  This 
is already in progress through, e.g., the Nearby Supernova Factory 
and Carnegie Supernova Project. 

\item Testing of astrophysical systematic effects on supernova 
distances, e.g.\ theoretical studies of gravitational lensing 
magnification, observational studies of dust extinction properties, 
multicolor flux calibration.  One example of projects in progress 
with HST is the ``Decelerating and Dustfree'' program, studying $z>1$ 
supernovae in clusters of elliptical galaxies where extinction corrections 
should be minimal.  The Canada-France-Hawaii Telescope Supernova 
Legacy Survey and CTIO Essence Project are 
underway to provide detailed multicolor light curves for 
over 800 Type Ia supernovae. 

\item Weak lensing robustness in both larger observational data sets 
and improved algorithmic treatment of extracting the signal and 
separating telescope, atmosphere, and intrinsic systematics.  
The CFHT Legacy Survey is in progress to deliver some 140 square 
degrees of data to moderate depth and galaxy number density.  
Pan-Starrs is gearing up for a two order of magnitude increase 
over this area, within the next five years.  On the analysis side, 
a widely international collaboration is testing data extraction 
through the Shear Testing Programs (STEP). 

\end{enumerate} 

These requirements to ``ensure rapid progress'' in understanding dark 
energy are already underway.  Dark energy is such a fundamental question 
that the community has not sat back and waited for this generation of 
experiments, but moved forward to make it happen.  Nor should the 
revelatory and robust experiments of the succeeding generation, JDEM 
and LST, remain ``in the dugout''.  They should be on deck, warming up, 
with the precursor lead-off experiments setting up the conditions for 
their home run on dark energy. 

The two key components of this strategy not already underway are: 
1) a comprehensive astrophysical flux calibration program, and 2) a 
comprehensive cosmological theory program so that interpretation of 
the incoming and forthcoming data will not be theory limited. 
Note that both components will have wide impact throughout 
astrophysics and cosmology while at the same time being the 
critical precursors to understanding dark energy. 

Other hopes exist for methods of probing our universe; for example 
baryon acoustic oscillations appears promising.  If 
baryon oscillation surveys achieve their potential they can help 
supernovae in distance determination; note, however, that baryon 
oscillations alone, even at high precision on distances, cannot 
match the cosmological parameter leverage of supernova distances, 
nor is baryon oscillations plus weak lensing the equal 
of supernovae plus weak lensing.  This arises because baryon oscillations 
give distances relative to high redshift, where dark energy is 
negligible, rather than to low redshift where dark energy is dominant 
\citep{linbo,boup}. 
But baryon oscillations will have 
a different, and possibly benign, set of systematics -- issues 
of mode coupling, 
redshift space distortions, biasing, and simply obtaining sufficient 
high S/N observations need to be actively researched. 
With finite resources, decisions need to be made to ensure precursor 
projects should not delay revelatory, robust results for 
any longer than needed in direct support of that goal. 

\section{Overview of SNAP} 

The data characteristics for JDEM, as SNAP, that are driven 
by the science requirements for a revelatory and robust dark energy 
experiment are: 

\begin{itemize} 
\item Full redshift range $z=0-1.7$ with dense sampling, to 
break parameter degeneracies and bound systematics. 
\item $\sim$2000 supernovae with optical/near infrared imaging 
and spectra to 1) divide into subsets for like-to-like comparison 
(``anti-evolution''), 2) obtain high signal to noise to bound systematics 
and prevent Malmquist 
bias, and 3) obtain many $z>1$ supernovae to prevent gravitational 
lensing bias. 
\item Space telescope ($\sim$2 meter aperture) for 1) infrared 
observations (essential for high $z$) and high accuracy color (dust 
extinction) corrections, and 2) precise and stable weak gravitational 
lensing shear measurements. 
\item Crosschecking and complementary methods for robust 
characterization of the nature of dark energy.  Weak lensing adds 
great value to supernovae, in deep and wide surveys.  {\it No need 
for $\Omega_M$ prior!} 
\end{itemize} 

SNAP plans its observing strategy to maximize the science from both 
the supernova and weak lensing methods.  In the basic mission, the 
deep survey covers 15 square degrees repeatedly in 9 wavelength bands 
for 120 visits, discovering and following supernovae to $z\approx3$ 
and measuring lensing shears for $10^7$ 
galaxies with a number density of greater than 250 resolved galaxies 
per square arcminute.  This will be superb for a wide area dark matter 
map.  The basic wide survey scans 1000 square degrees 
once, down to AB 26.6 in each band, resolving 100 galaxies per square 
arcminute for a total of some 300 million galaxies.  With an 
extended mission, the wide survey can be expanded over additional 
thousands of square degrees. 

\subsection{SNAP Probes} 

These surveys automatically provide data for other dark energy 
methods.  For example SNAP will obtain a sample of some thousand 
clusters, measured in nine bands optically and through weak lensing, 
deeper in redshift and with a lower mass threshold 
than other proposed cluster surveys except for SPT. 
The wide survey maps baryon acoustic oscillations photometrically 
over 1000 square degrees (expandable) in the key redshift range 
$z\approx1-2$, problematic from the ground.  Such photometric 
surveys, while a factor 15 weaker than spectroscopic surveys for 
the same area, actually only require 4-5 times the area of a 
spectroscopic survey when used in complementarity with supernovae, 
to provide the same strength on the dark energy 
constraints. 

Note that space-based weak lensing (for which SNAP would establish 
{\it three orders of magnitude\/} improvement in area) provides 1) a 
higher density of 
resolved images, useful for probing smaller scale structure where 
the growth effects are amplified by nonlinearities, 2) deeper lenses 
allowing mapping of the mass growth over more cosmic time, 
3) accuracy allowing new kinds of weak lensing techniques such as 
cross-correlation cosmography, and 
4) elimination of systematics such as atmospheric distortion of the 
galaxy shapes and thermal, wind, and gravity loading of the telescope. 

\subsection{SNAP Complementarity} 

True synergy 
comes from bringing weak lensing and supernovae together.   In this case 
complementarity is achieved on several levels.  An experiment 
incorporating both techniques is truly comprehensive in that no 
external priors are required: no outside determination of the matter 
density is necessary.  Furthermore, 
the two methods conjoined provide a test of the spatial curvature of 
the universe to $\sim1-2\%$ (for the SNAP experiment), independent 
of the CMB constraint on flatness (note that the Planck CMB measurements 
in isolation would only determine the curvature to $\sim6\%$ 
\citep{eht}). 
On dark energy properties, supernovae plus weak lensing methods conjoined  
determine the present equation of state ratio, $w_0$, to 5\%, and 
its time variation, $w'$, to 0.11 (for the SNAP experiment basic 
mission, including an estimate of systematics, and in the relatively 
insensitive scenario of a true cosmological constant; this improves to 
$w_0$ to 0.03 and $w'$ to 0.06 in a fiducial SUGRA dark energy case).  
Finally, this 
synergy provides a real opportunity to test the physics framework, 
distinguishing between a new high energy physics component and an 
extension to Einstein gravity. 

SNAP, as an implementation of JDEM, is an experiment that can give a 
truly exciting view, revelatory and robust, into the nature of new 
fundamental physics.  No new technology or unproven methods are 
required.  In terms of a realistic technology and mission timeline, 
SNAP can be launched by 2012.  With the assistance of select, focused 
precursor experiments, the frontiers of science are within our reach. 

\vspace{0.2in} 

{\it Companion white papers give specifics on the SNAP supernova and 
weak lensing programs. 
For further, in-depth information see http://snap.lbl.gov and 
the comprehensive Aldering et al.\ article at 
http://arxiv.org/abs/astro-ph/0405232. }

\newpage 

\newpage 

\newpage 

\newpage 

\begin{thebibliography} 

\bibitem[Caldwell \& Linder(2005)]{caldlin} 
 R.R.\ Caldwell \& E.V.\ Linder 2005, {\it The 
Limits of Quintessence\/}, 
submitted to Phys.\ Rev.\ Lett.\ [astro-ph/0505494] 

\bibitem[Eisenstein, Hu, \& Tegmark(1999)]{eht} 
D.J.\ Eisenstein, W.\ Hu, M.\ Tegmark 1999, 
{\it Cosmic Complementarity: Joint Parameter Estimation from Cosmic Microwave Background Experiments and Redshift Surveys\/}, Ap.\ J.\ 518, 2 
[astro-ph/9807130] 

\bibitem[Linder(2003)]{linbo} 
E.V.\ Linder 2003, {\it Baryon Oscillations as a Cosmological Probe\/}, 
Phys.\ Rev.\ D 68, 083504 [astro-ph/0304001] 

\bibitem[Linder(2005a)]{lin05a} 
E.V.\ Linder 2005a, {\it Cosmic Growth History and Expansion History\/}, 
astro-ph/0507263 

\bibitem[Linder(2005b)]{boup} 
E.V.\ Linder 2005b, {\it The Ups and Downs of Baryon Oscillations\/}, 
astro-ph/0507308

\bibitem[Linder \& Huterer(2003)]{linhut03} 
E.V.\ Linder \& D.\ Huterer 2003, {\it Importance of Supernovae at $z>1.5$  
to Probe Dark Energy\/}, Phys.\ Rev.\ D 67, 081303 [astro-ph/0208138] 

\bibitem[Linder \& Miquel(2004)]{linmiq} 
E.V.\ Linder \& R.\ Miquel 2004, {\it Is Dark Energy Dynamical? Prospects for 
an Answer\/}, Phys.\ Rev.\ D 70, 123516 [astro-ph/0409411] 

\bibitem[Vale \& Linder(2004)]{valelin} 
C.\ Vale \& E.V.\ Linder 2004, in {\it The Darkness of the Universe: 
Mapping Expansion and Growth\/}, http://supernova.lbl.gov/\~{}evlinder/taiwan3.ppt 

\bibitem[Wood-Vasey et al.(2004)]{woodvasey} 
W.M.\ Wood-Vasey et al.\ 2004, {\it The Nearby Supernova Factory\/}, 
New Astron.\ Rev.\ 48, 637 [astro-ph/0401513] 

\end{thebibliography}
\end{document}